\documentclass[lettersize,journal]{IEEEtran}
\usepackage{amsmath,amsfonts}
\usepackage{array}
\usepackage[caption=false,font=normalsize,labelfont=sf,textfont=sf]{subfig}
\usepackage{textcomp}
\usepackage{stfloats}
\usepackage{url}
\usepackage{verbatim}
\usepackage{graphicx}
\usepackage{cite}
\usepackage{xcolor}
\usepackage{eso-pic} 
\usepackage{booktabs}
\usepackage{multirow}
\usepackage{makecell}
\usepackage{hyperref}

\usepackage[ruled,vlined]{algorithm2e}
\SetKw{Continue}{continue}
\SetKw{Break}{break}

\begin{document}

\title{ChroNoC: A Tableless TDM Circuit-Switched NoC With Delay-Aware Deployment for Time-Sensitive Neuromorphic Computing}

\author{Mohammad~Rasoul Roshanshah~\textsuperscript{\href{https://orcid.org/0000-0001-5650-7925}{\includegraphics[width=8pt]{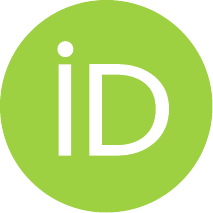}}}, 
Mehdi Modarressi~\textsuperscript{\href{https://orcid.org/0000-0002-4117-7609}{\includegraphics[width=8pt]{orcid.pdf}}},
and~Saeed Safari~\textsuperscript{\href{https://orcid.org/0000-0001-6940-591X}{\includegraphics[width=8pt]{orcid.pdf}}}
\thanks{The authors are with the School of Electrical and Computer Engineering, College of Engineering, University of Tehran, Tehran, Iran.} \thanks{(e-mails: mrroshanshah@ut.ac.ir; modarressi@ut.ac.ir; saeed@ut.ac.ir)}
}

\maketitle

\AddToShipoutPictureFG*{%
  \AtTextLowerLeft{%
    \put(\LenToUnit{0.5\textwidth},-8){%
      \makebox(0,0)[t]{%
        \begingroup
        \setlength{\fboxrule}{0.6pt}%
        \setlength{\fboxsep}{3pt}%
        \fcolorbox{black}{white}{%
          \begin{minipage}{0.80\textwidth}
            \normalfont\fontsize{8}{9.5}\selectfont
            \setlength{\parindent}{0pt}%
            \setlength{\parskip}{0pt}%
            This work has been submitted to the IEEE for possible publication.
            Copyright may be transferred without notice, after which this version
            may no longer be accessible.
          \end{minipage}%
        }%
        \endgroup
      }%
    }%
  }%
}
	
\begin{abstract}
Time-sensitive neuromorphic architectures require predictable interconnects to preserve precise event timing. 
Statically scheduled time-division-multiplexed (TDM) circuit-switched networks-on-chip (NoCs) provide deterministic latency at the cost of linear configuration-memory scaling with schedule depth. 
Furthermore, jointly optimizing task placement, routing, and scheduling creates an intractable combinatorial search space. 
Existing deployment frameworks also ignore algorithmic axonal delays by forcing immediate spike injections that increase network contention. 
This paper proposes a hardware-software co-design to address these challenges. 
ChroNoC is a tableless TDM NoC that embeds precomputed full-grid broadcast trees into combinational logic as a fixed routing substrate. 
This fixed substrate reduces deployment to a spatial search via Delay-Aware Reservations and Spatial Annealing (DARSA). 
DARSA converts axonal delays into periodic scheduling slack to stagger network injections and reduce communication overhead. 
In a 130\,nm CMOS process, ChroNoC reduces per-hop energy consumption by more than 75\% relative to the evaluated circuit- and packet-switched baselines. 
On UCI-HAR and N-MNIST workloads, DARSA reduces per-timestep communication cycles by approximately 60\% and 45\%, respectively, compared with immediate transmission, while also reducing provisioned communication memory. 
These results demonstrate how combining lightweight TDM switch design with delay-aware deployment improves both hardware efficiency and system-level communication performance while preserving neural event timing.
\end{abstract}
	

\section{Introduction}
\label{sec:introduction}

\IEEEPARstart{T}{ime}-sensitive many-core systems require predictable communication under concurrent traffic \cite{valente2024top}.
Network-on-Chip (NoC) architectures provide scalable on-chip connectivity, but contention, queueing, and arbitration can make packet delivery depend on competing traffic \cite{picornell2021dcfnoc,dally2001route,benini2002networks,shi2008real}.
Circuit switching reserves communication resources in advance \cite{goossens2005aethereal,liu2013analysis}.
A dedicated circuit may nevertheless leave links idle between events and restrict other transfers that require the same resources.
Time-Division Multiplexing (TDM) shares these resources through recurring timeslots \cite{hansson2007channel,lu2008tdm}.
Platforms such as dAElite \cite{stefan2014daelite}, Argo \cite{kasapaki2016argo}, and T-CREST \cite{schoeberl2015tcrest} demonstrate the value of statically scheduled communication for predictable execution.

Conventional table-based TDM fabrics store the switching configuration for each timeslot \cite{roshanshah2026chronospike,stefan2014daelite,sparso2022comparing}. This storage grows linearly with schedule depth.
A fixed schedule permits another implementation.
The crossbar controls can be synthesized directly as a function of the timeslot counter.
This removes programmable per-timeslot configuration memory.

Predictable event delivery is also relevant to neuromorphic computing, where distributed neurons exchange spikes \cite{mead1990neuromorphic,roy2019towards}.
Recent advances in temporal coding and delay learning offer a promising basis for lightweight spiking models targeting resource-constrained edge applications.
Single-spike coding represents information through firing times with limited neuronal activity \cite{moqadasi2025temporal}, while trainable transmission delays enrich temporal processing \cite{sun2023learnableaxonal,dagostino2024denram}.
DelGrad \cite{goltz2025delgrad} demonstrates this combination in single-spike models, showing that jointly training weights and delays can improve accuracy and parameter efficiency on neuromorphic hardware.
Once trained and fixed for inference, these delays define permissible delivery intervals, while single-spike execution bounds source event counts.
Together, these properties enable communication provisioning that exploits temporal flexibility instead of requiring immediate transmission of every spike.

Existing neuromorphic mappers focus primarily on objectives such as spike volume, hop count, congestion, and locality~\cite{balaji2020spinemap,li2020sneap,jin2025alp,urgese2016optimizing}. 
Although prior work recognizes prescribed delays as a hardware resource~\cite{kim2018axonal}, state-of-the-art tools enforce delays only after delivery rather than exploiting them as injection scheduling slack. 
Without this temporal flexibility, spikes released at the same timestep concentrate into narrow injection windows that inflate the physical frame duration. 
Conversely, delivering events far ahead of their consumption time forces destination cores to allocate excessive inter-timestep buffering \cite{akopyan2015truenorth,rhodes2018spynnaker}. 

Addressing these bottlenecks requires coordinating neuron placement, physical routing, and permissible delivery windows. 
However, jointly assigning logical tasks $\mathcal{A}$ to physical cores $\mathcal{V}$ while selecting among $R_{\max}$ candidate routes and $\mathcal{T}$ TDM injection slots for each communication demand $\mathcal{D}$ yields a combinatorial search space bounded by 
\begin{equation}
    |\Omega_{\mathrm{joint}}|
    \leq
    |\mathcal{V}|^{|\mathcal{A}|}
    R_{\max}^{|\mathcal{D}|}
    \mathcal{T}^{|\mathcal{D}|}.
    \label{eq:joint_complexity}
\end{equation} 
To eliminate the $R_{\max}^{|\mathcal{D}|}\mathcal{T}^{|\mathcal{D}|}$ factor, this paper decouples physical schedule construction from workload deployment.
ChroNoC provides one fixed space-time broadcast tree per source core using counter-driven synthesized crossbar decoding. 
Assigning a neuron to a core automatically fixes its broadcast tree and within-frame injection slot, removing both decisions from the deployment search. 

On this fixed substrate, Delay-Aware Reservations and Spatial Annealing (DARSA) searches capacity-constrained neuron placements while deterministically deriving each candidate's periodic calendar from the prescribed delays. 
Excluding independent route and slot variables reduces the deployment search space to
\begin{equation}
    |\Omega_{\mathrm{fixed}}|
    \leq
    |\mathcal{V}|^{|\mathcal{A}|}.
    \label{eq:fixed_complexity}
\end{equation}
Although restricting solutions to a fixed broadcast substrate and deterministic calendar family sacrifices unrestricted joint optimality, it reduces deployment to a spatial placement problem addressed using simulated annealing.

The main contributions are summarized below. 
\begin{enumerate}
    \item \textbf{Tableless deterministic multicast.}
    ChroNoC implements fixed, conflict-free broadcast trees using synthesized crossbar-control logic, eliminating programmable per-timeslot switch memories and supporting multicast with one network injection per source event.

    \item \textbf{Decoupled deployment optimization.}
    A reusable physical routing substrate removes independent route and within-frame injection-slot decisions from the deployment search. DARSA optimizes capacity-constrained neuron placement while balancing frame duration and provisioned communication memory.

    \item \textbf{Deadline-preserving periodic reservations.}
    DARSA exploits prescribed axonal delays as scheduling slack to stagger network injections without advance knowledge of firing times. Under the stated single-spike execution assumptions, these reservations preserve logical event-consumption times.
\end{enumerate}

\section{Background and Motivation}
\label{sec:background}

Spiking Neural Networks (SNNs) use the timing of discrete spikes to process information \cite{eissa2025stems,jannesar2024sste}.
The sparsity of these spikes can also be exploited to reduce energy consumption in computation and communication \cite{fereidani2027rescom}.
Distributed implementations execute these networks by exchanging discrete spikes across many-core architectures \cite{furber2013overview,davies2018loihi,su2024ultralowcost,akopyan2015truenorth}.
To communicate across these distributed cores, hardware architectures predominantly rely on the Address-Event Representation (AER) protocol \cite{boahen2000point,furber2014spinnaker}.
In AER, biological action potentials are abstracted into lightweight digital packets containing the unique address of the firing source neuron \cite{farahani2026flexineura}.

Because each neuron typically projects to multiple destinations, hardware multicast is used to replicate an injected packet across network branches \cite{su2024ultralowcost,merolla2014multicast,ghavami2026suprasnn}. 
Even in temporally sparse SNNs, neurons that fire during the same timestep generate heavy multicast traffic and severe injection contention. 

Two algorithmic properties of modern SNNs offer an opportunity to relieve this contention. 
First, single-spike coding restricts each neuron to at most one spike per input sample \cite{moqadasi2025temporal,goltz2025delgrad}. 
Second, delay-learning frameworks such as DelGrad train synaptic weights alongside transmission delays to improve model accuracy and parameter efficiency \cite{goltz2025delgrad,sun2023learnableaxonal}.
In particular, an axonal delay assigns a single delay to all outgoing connections of a source neuron.
These axonal delays are fixed prior to hardware deployment.
During operation, input samples are processed sequentially after all spikes from the previous inference have been consumed.
Although the exact firing timestep of each neuron remains unknown until runtime, single-spike coding bounds the total number of generated events \cite{moqadasi2025temporal,goltz2025delgrad}.
At the same time, the fixed axonal delay defines a permissible delivery window for every spike.

To preserve exact neural behavior on a statically scheduled TDM interconnect, physical network arrival must be distinguished from logical synaptic consumption.
In a time-stepped SNN \cite{davies2018loihi,rhodes2018spynnaker}, each algorithmic timestep updates neuron states before a TDM communication frame transfers spikes across the fabric. 
Suppose source neuron $s$ fires at timestep $g_s$ prior to communication frame $g_s$.
Its axonal delay $\delta_s^{\mathrm{ax}}\geq1$ sets the logical consumption deadline $h_s=g_s+\delta_s^{\mathrm{ax}}$ across all destination cores.
A spike injected during frame $f_s$ completes physical delivery by the end of that frame at boundary $f_s+1$.
Preserving the reference neural timing therefore requires 
\begin{equation}
    g_s\leq f_s,
    \qquad
    f_s+1\leq h_s
    =g_s+\delta_s^{\mathrm{ax}}.
    \label{eq:deadline}
\end{equation}
Any spike that arrives before its consumption deadline $h_s$ is held in a destination buffer until that timestep is reached \cite{akopyan2015truenorth}.
For instance, SpiNNaker \cite{furber2014spinnaker} relies on dedicated buffers to manage these early arrivals \cite{rhodes2018spynnaker}.

Existing deployment tools ignore the scheduling slack in (\ref{eq:deadline}) by forcing every spike to be injected immediately in frame $f_s=g_s$. 
Immediate injection requires worst-case per-frame bandwidth to accommodate simultaneous firing while forcing destination cores to buffer long-delay spikes across many timesteps. 
Instead, an axonal delay of $\delta_s^{\mathrm{ax}}$ timesteps allows network injection to be spread across up to $\delta_s^{\mathrm{ax}}$ consecutive frames without predicting firing times or altering neural states. 
Staggering injections across frames directly reduces the peak per-frame injection capacity required by each core. 
Because the spatial placement of neurons determines both the number of sources and the minimum axonal delay on each core, placement directly controls local reservation periods, global TDM frame duration, and buffer storage. 
This tight coupling between spatial placement and temporal delay slack motivates the proposed co-design of a fixed broadcast fabric and a delay-aware deployment framework.

\section{Architecture and Deployment}
\label{sec:architecture}

\begin{figure*}[t]
    \centering
    \includegraphics[width=\linewidth]{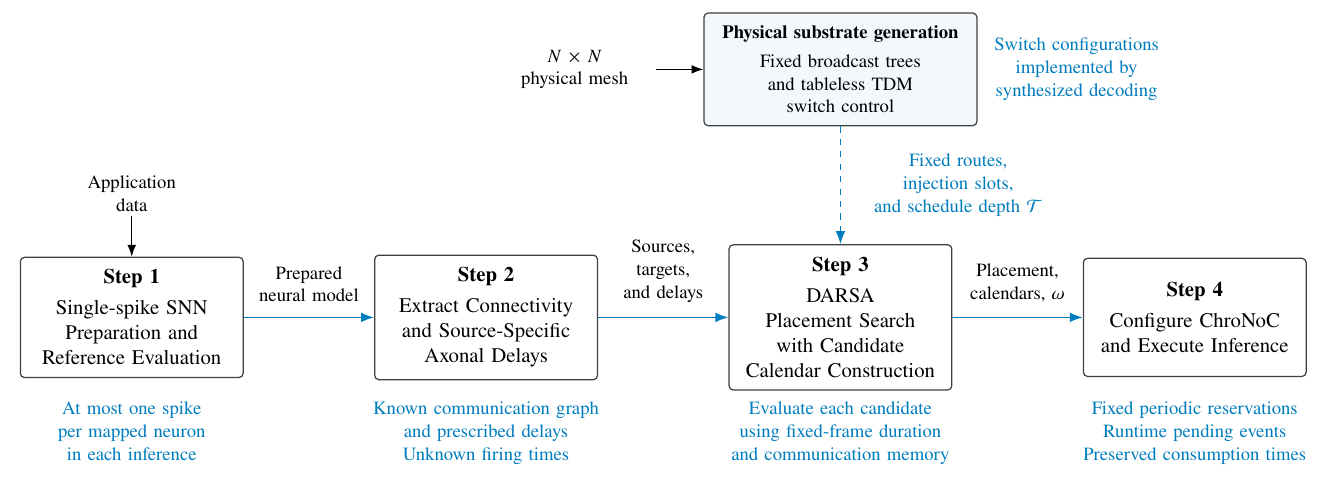}
    \caption{Workload preparation and deployment on ChroNoC using workload-independent broadcast routing and delay-aware periodic reservations.} 
    \label{fig:deployment_flow}
\end{figure*}

The deployment flow decouples physical routing, neuron placement, and source reservation assignment (Fig.~\ref{fig:deployment_flow}). 
Software evaluation first extracts workload connectivity and source-specific axonal delays without requiring runtime firing times. 
Independently, offline routing generates reusable broadcast trees, core injection timeslots, and switch configurations from the mesh topology. 
During placement search, DARSA constructs a deterministic Network Interface (NI) calendar for each candidate to evaluate frame duration and provisioned memory. 
After configuring the selected placement and calendars, each NI injects pending runtime events during its reserved opportunities for deterministic delivery and deadline-accurate consumption. 

\subsection{Fixed Broadcast Substrate}
\label{subsec:fixed_broadcast_substrate}

\begin{figure}[t]
    \centering
    \includegraphics[width=0.95\linewidth]{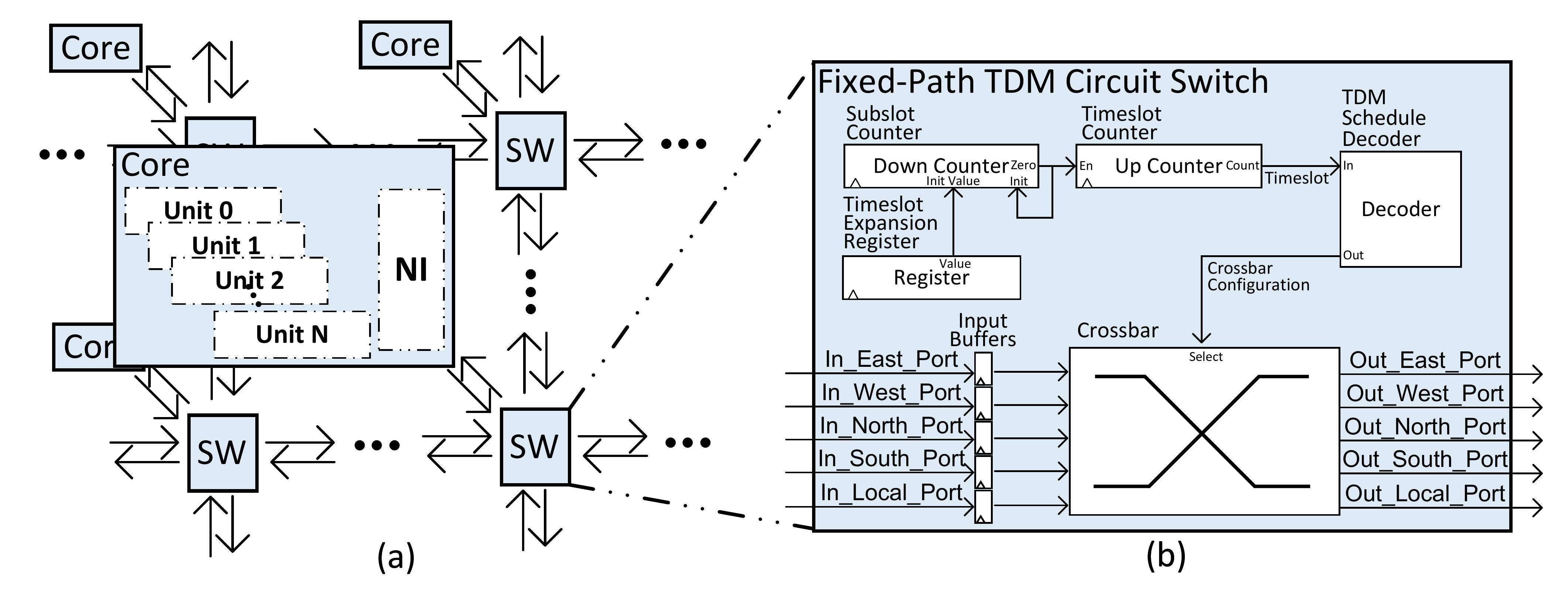}
    \caption{ChroNoC mesh and tableless TDM switch using counter-driven synthesized crossbar decoding.} 
    \label{fig:architecture}
\end{figure}

ChroNoC interconnects cores via a two-dimensional mesh of circuit switches with one full-grid broadcast tree per source core (Fig.~\ref{fig:architecture}). 
Because circuit paths are pre-established without routing headers, each transmitted AER spike word is referred to as a datagram. 
Injected datagrams replicate along fixed tree branches independent of neural connectivity, leaving receiving cores to filter for local target neurons. 
Offline routing fixes each core's injection timeslot and every per-timeslot switch configuration for reuse across placements. 
Inside each switch, a synchronized timeslot counter drives synthesized combinational decoding to establish one-to-many crossbar connections. 
This tableless design eliminates per-timeslot configuration memories and runtime header routing, retaining only the counter, decoder logic, and crossbar wiring. 

\subsection{Offline Space-Time Broadcast Routing}
\label{subsec:offline_broadcast_routing}

\begin{figure}[t!]
    \centering
    \centerline{\includegraphics[width=0.6\linewidth]{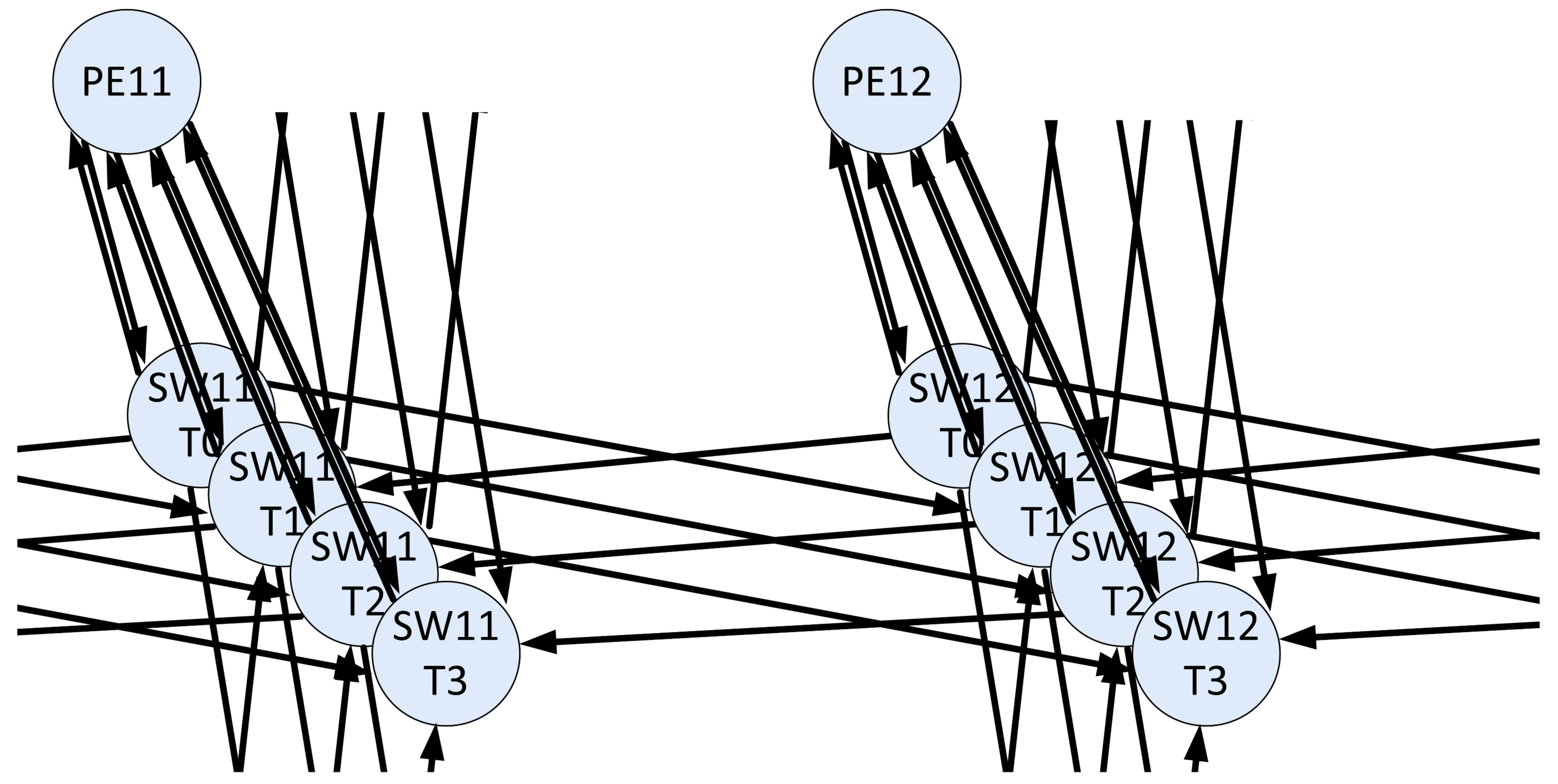}}
    \caption{Time-expanded graph construction.} 
    \label{fig:st_graph}
\end{figure}

For an $N\times N$ mesh, the offline router constructs $N^2$ workload-independent all-to-all broadcast trees $\mathcal{R}$, core injection timeslots $\boldsymbol{\Phi}$, arrival times $\mathbf{a}$, and schedule depth $\mathcal{T}$. 
Directed core-to-switch and inter-switch links are unfolded (Fig.~\ref{fig:st_graph}) over a search horizon of $H=4N^2+4N$ logical timeslots into an acyclic graph $\mathcal{G}_H$ with vertices $(u,\ell)$ and unit-step edges 
\begin{equation}
    (u,\ell)\rightarrow(v,\ell+1),
    \qquad 0\leq\ell<H-1.
    \label{eq:space_time_edge}
\end{equation} 
We use a search horizon of $H=4N^2+4N$ timeslots and retain only validated, conflict-free schedules.
Reserving link-timeslot pairs ensures conflict-free routing across repeated link traversals. 

Algorithm~\ref{alg:main-workflow} executes $I_{\max}$ randomized routing attempts from empty reservation sets. 
Each attempt shuffles the core order and samples each injection slot $\Phi_v$ according to 
\begin{equation}
    \Phi_v\in
    \left\{0,\ldots,\max(1,H-2N-2)-1\right\}.
    \label{eq:routing_start_range}
\end{equation} 
The injection-slot range leaves a within-frame propagation margin; route validation checks that all datagrams reach their destinations within the search horizon $H$.
Unlike the inter-frame delayed-event tail at the end of an inference, this within-frame propagation tail is embedded within $\mathcal{T}$ and scales by $\omega$ during subslot expansion to drain all in-flight subslots before each frame boundary. 
The broadcast tree for core $v$ grows incrementally from $(v,\Phi_v)$ to all other cores via multi-source breadth-first search (BFS) initialized with $(v,\Phi_v)$ and current tree switch vertices. 
Uniform edge costs allow BFS to add shortest available branches while reusing existing tree segments. 
Cores act strictly as launch roots or terminal endpoints rather than transit or waiting nodes. 
Each search excludes previously reserved link-timeslots and locks the new tree's resources upon completion.

\begin{algorithm}[t]
    \caption{Offline Space-Time Broadcast Routing}
    \label{alg:main-workflow}
    \KwIn{
        Physical mesh $\mathcal{G}$ with core set $\mathcal{V}$;
        search horizon $H$;
        routing attempts $I_{\max}$
    }
    \KwOut{
        Best-found broadcast trees $\mathcal{R}$;
        injection slots $\boldsymbol{\Phi}$;
        arrival times $\mathbf{a}$;
        schedule depth $\mathcal{T}$
    }

    $\mathcal{G}_H
        \gets \textsc{UnfoldMesh}(\mathcal{G},H)$\;
    $\mathcal{R}_{\mathrm{best}}\gets\emptyset$;
    $\mathcal{T}_{\mathrm{best}}\gets+\infty$\;

    \For{$i\gets1$ \KwTo $I_{\max}$}{
        $\mathcal{U}\gets\emptyset$;
        $\mathcal{R}\gets\emptyset$;
        $success\gets\mathrm{true}$\;

        \ForEach{$v$ in $\textsc{Shuffle}(\mathcal{V})$}{
            $\Phi_v\gets
                \textsc{SampleInjectionSlot}(H,N)$\;
            $\mathcal{R}_v\gets\emptyset$\;
            $\mathcal{W}_v\gets\{(v,\Phi_v)\}$\;
            $\mathcal{Q}\gets\mathcal{V}\setminus\{v\}$\;
            $a_{v,v}\gets\Phi_v$\;

            \While{$\mathcal{Q}\neq\emptyset$}{
                $\mathcal{B}\gets
                    \{(v,\Phi_v)\}
                    \cup
                    \textsc{SwitchVertices}(\mathcal{W}_v)$\;

                $p\gets
                    \textsc{MultiSourceBFS}
                    (\mathcal{G}_H,\mathcal{B},
                     \mathcal{Q},\mathcal{U})$\;

                \If{$p=\bot$}{
                    $success\gets\mathrm{false}$\;
                    \Break\;
                }

                $(u,\ell)\gets\textsc{TerminalVertex}(p)$\;
                $\mathcal{R}_v\gets
                    \mathcal{R}_v\cup\textsc{Edges}(p)$\;
                $\mathcal{W}_v\gets
                    \mathcal{W}_v\cup\textsc{Vertices}(p)$\;
                $a_{v,u}\gets\ell$\;
                $\mathcal{Q}\gets\mathcal{Q}\setminus\{u\}$\;
            }

            \If{\textnormal{not} $success$}{
                \Break\;
            }

            $\mathcal{U}\gets\mathcal{U}\cup\mathcal{R}_v$\;
            $\mathcal{R}\gets
                \mathcal{R}\cup\{v\mapsto\mathcal{R}_v\}$\;
        }

        \If{$success$}{
            $\mathcal{T}\gets1+\max_{v,u}a_{v,u}$\;
            $\textsc{ValidateRoutes}
                (\mathcal{R},\boldsymbol{\Phi},
                 \mathbf{a},\mathcal{T})$\;

            \If{$\mathcal{T}<\mathcal{T}_{\mathrm{best}}$}{
                Save $(\mathcal{R},\boldsymbol{\Phi},
                       \mathbf{a},\mathcal{T})$
                as the best schedule\;
            }
        }
    }

    \If{$\mathcal{R}_{\mathrm{best}}=\emptyset$}{
        \KwRet routing failure\;
    }
    \KwRet best schedule\;
\end{algorithm}

Failed attempts restart from an empty reservation set. 
Successful attempts yield schedule depth $\mathcal{T}=1+\max_{v,u} a_{v,u}$. 
The smallest-depth validated schedule is retained and synthesized offline into the tableless crossbar decoders.

\subsection{Periodic Injection Reservations}
\label{subsec:periodic_reservations}

Each core $v$ hosts $m_v=|\mathcal{S}_v|\le C_v$ transmitting sources. 
Without runtime firing predictions, the NI calendar assigns each source a recurring frame phase and injection subslot. 
Sources on a nonempty core $v$ share a common reservation period $P_v$ and maximum per-phase reservation count $c_v$ 
\begin{equation}
    P_v=\min_{s\in\mathcal{S}_v}\delta_s^{\mathrm{ax}},
    \qquad
    c_v=\left\lceil\frac{m_v}{P_v}\right\rceil,
    \label{eq:period}
\end{equation} 
with $P_v=1$ and $c_v=0$ when $\mathcal{S}_v=\emptyset$. 
Ordering sources by index $j\ge 0$ distributes them across phases according to 
\begin{equation}
    \theta_s=j\bmod P_v,
    \label{eq:calendar_assignment}
\end{equation} 
where sources sharing the same phase $\theta_s$ occupy consecutive injection subslots starting from zero. 
The global expansion factor and frame length are 
\begin{equation}
    \omega=\max\left\{1,\max_v c_v\right\},
    \qquad
    L_{\mathrm{frame}}=\mathcal{T}\omega.
    \label{eq:capacity}
\end{equation} 

An event generated at timestep $g_s$ boards the first eligible frame 
\begin{equation}
    f_s=g_s+
    \bigl((\theta_s-g_s)\bmod P_v\bigr).
    \label{eq:boarding}
\end{equation} 
Because waiting never exceeds $P_v-1$ frames, delivery completes by the consumption deadline $h_s$ regardless of firing time 
\begin{equation}
    f_s+1
    \leq g_s+P_v
    \leq g_s+\delta_s^{\mathrm{ax}}
    =h_s.
    \label{eq:guarantee}
\end{equation} 
Therefore, the core period $P_v$ governs transmission repetition without altering individual consumption deadlines $h_s=g_s+\delta_s^{\mathrm{ax}}$.

\subsection{Communication Memory Accounting}
\label{subsec:communication_memory}

While sources buffer pending events until injection, destinations retain early arrivals for the logical waiting interval 
\begin{equation}
    W_s
    =
    h_s-(f_s+1)
    =
    g_s+\delta_s^{\mathrm{ax}}-(f_s+1)
    \geq0.
    \label{eq:retention_interval}
\end{equation} 
Conservative provisioning for unknown firing times allocates one pending entry per transmitting source and one shared destination entry per source--target core pair prior to local synaptic expansion. 
For source $s$ with target neurons $\mathcal{D}_s$ under placement $\mathcal{M}$, the target-core indicator (including local-core targets) is 
\begin{equation}
    y_{s,v}
    =
    \mathbb{I}
    \left[
        \exists d\in\mathcal{D}_s
        \text{ with }\mathcal{M}(d)=v
    \right].
    \label{eq:target_core_indicator}
\end{equation} 
With event-entry width $b_{\mathrm{evt}}$ (source ID, timing tag, valid bit), source and destination provisions are 
\begin{align}
    B_{\mathrm{src}}
        &=|\mathcal{S}_{\mathrm{all}}|b_{\mathrm{evt}},
        \label{eq:source_memory}\\
    B_{\mathrm{dst}}
        &=b_{\mathrm{evt}}
          \sum_{s\in\mathcal{S}_{\mathrm{all}}}
          \sum_{v\in\mathcal{V}}y_{s,v}.
        \label{eq:destination_memory}
\end{align} 
Total provisioned communication memory is 
\begin{equation}
    B_{\mathrm{tot}}
    =
    B_{\mathrm{src}}
    +
    B_{\mathrm{dst}}
    +
    B_{\mathrm{cal}}
    +
    B_{\mathrm{NI}}
    +
    B_{\mathrm{net}},
    \label{eq:total_communication_memory}
\end{equation} 
combining phase/subslot counters and $P_v c_v$ calendar entries per core ($B_{\mathrm{cal}}$), NI datapath registers ($B_{\mathrm{NI}}$), and two $\omega$-word banks per receiving channel ($B_{\mathrm{net}}$). 
Neural states, weights, and trace-specific runtime queue peaks are excluded. 

\subsection{Spatial Placement}
\label{subsec:DARSA_mapping}

DARSA maps logical neurons $\mathcal{A}$ onto physical cores $\mathcal{V}$ over the fixed substrate $(\mathcal{R},\boldsymbol{\Phi},\mathcal{T})$ using the binary variable 
\begin{equation}
    x_{a,v}
    =
    \begin{cases}
        1, & \mathcal{M}(a)=v,\\
        0, & \text{otherwise},
    \end{cases}
    \label{eq:mapping_variable}
\end{equation} 
subject to unique assignment and core task capacity $C_v$ (distinct from the per-phase reservation count $c_v$) 
\begin{align}
    \sum_{v\in\mathcal{V}}x_{a,v}
        &=1,
        &&\forall a\in\mathcal{A},
        \label{const:unique_map}\\
    \sum_{a\in\mathcal{A}}x_{a,v}
        &\leq C_v,
        &&\forall v\in\mathcal{V}.
        \label{const:cap}
\end{align} 
Each candidate $\mathcal{M}$ deterministically defines $\mathcal{S}_v$ and calendar $\mathcal{C}(\mathcal{M})$ via (\ref{eq:period})--(\ref{eq:capacity}) without firing traces. 
DARSA minimizes the equally weighted ($w_L=w_B=0.5$) objective 
\begin{equation}
    F(\mathcal{M})
    =
    w_L
    \frac{\mathcal{T}\omega(\mathcal{M})}{L_0}
    +
    w_B
    \frac{B_{\mathrm{tot}}(\mathcal{M})}{B_0},
    \label{eq:objective}
\end{equation} 
normalized by the initial frame duration $L_0$ and memory $B_0$ of the cyclic Round-Robin baseline 
\begin{equation}
    \mathcal{M}_0(a_j)
    =
    j\bmod|\mathcal{V}|.
    \label{eq:round_robin_initialization}
\end{equation} 

DARSA's annealing algorithm explores placements via equally probable single-neuron moves (skipping cost evaluation if the target core is full) and two-neuron swaps. 
Valid candidates rebuild $\mathcal{C}(\mathcal{M}')$ and are accepted if $F(\mathcal{M}')\le F(\mathcal{M})$ or with uphill probability 
\begin{equation}
    p_{\mathrm{accept}}
    =
    \exp\left(
        -\frac{F(\mathcal{M}')-F(\mathcal{M})}{\tau_k}
    \right),
    \qquad
    F(\mathcal{M}')>F(\mathcal{M}),
    \label{eq:annealing_acceptance}
\end{equation} 
under geometric cooling ($\tau_0=0.15$, $\tau_{\mathrm{end}}=0.001$, distinct from $\mathcal{T}$) over $K$ attempts or time budget $B_{\mathrm{time}}$ 
\begin{equation}
    \tau_k
    =
    \tau_0
    \left(
        \frac{\tau_{\mathrm{end}}}{\tau_0}
    \right)^{k/\max(1,K-1)},
    \qquad k=0,\ldots,K-1.
    \label{eq:annealing_temperature}
\end{equation}

\section{Experimental Results}
\label{sec:results}

This section evaluates the proposed architecture and deployment framework through hardware implementation and deployment experiments.

\subsection{Hardware Implementation and Comparative Efficiency}
\label{subsec:hardware_implementation_efficiency}

ChroNoC was implemented in Verilog as a $4\times4$ mesh of 16 switches with 32-bit unidirectional datapaths and a schedule depth of $\mathcal{T}=64$.
Physical implementation used the OpenLane RTL-to-GDSII flow~\cite{shalan2020building} with the open-source SkyWater 130\,nm CMOS Process Design Kit (PDK)~\cite{sky130}. 
Table~\ref{tab:baseline_metrics} summarizes the post-layout results. 

\begin{table}[t]
    \centering
    \caption{ChroNoC baseline post-layout characterization.}
    \label{tab:baseline_metrics}
    \begin{tabular}{lr}
        \toprule
        \textbf{Parameter / metric} & \textbf{Value} \\
        \midrule
        Topology / switches & $4\times4$ mesh / 16 \\
        Process technology & 130\,nm \\
        Clock frequency & 100\,MHz \\
        Per-direction link width & 32 bits \\
        Schedule depth ($\mathcal{T}$) & 64 \\
        Die dimensions & $1.6\times1.6$\,mm \\
        Core area & 2.5066\,mm$^2$ \\
        Total power & 26.1138\,mW \\
        \bottomrule
    \end{tabular}
\end{table}

\begin{figure}[!t]
    \centering
    \includegraphics[width=0.95\linewidth]{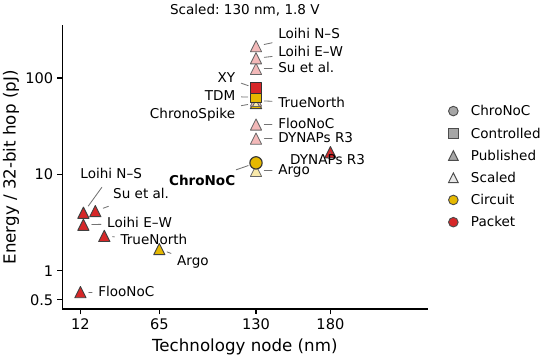}
    \caption{Energy per hop across technology nodes (logarithmic axis).} 
    \label{fig:energy_comparison}
\end{figure}

Compared against a table-based TDM switch and a four-stage XY wormhole router at 1.8\,V and 100\,MHz, the ChroNoC switch reduces hop energy by 79.25\% and 83.14\%, respectively (Fig.~\ref{fig:energy_comparison}). 
These savings result from eliminating routing/switching tables, runtime arbitration, and deep buffers. 

Figure~\ref{fig:energy_comparison} also compares ChroNoC with published packet-switched, neuromorphic, and circuit-switched routers. 
Low-energy packet-switched designs such as FlooNoC~\cite{fischer2025floonoc}, Su et al.~\cite{su2024ultralowcost}, Loihi~\cite{davies2018loihi}, DYNAPs R3~\cite{moradi2017scalable}, and TrueNorth~\cite{akopyan2015truenorth} rely on runtime arbitration, buffering, or asynchronous handshaking. 
Consequently, they incur traffic-dependent latency. 
Among deterministic designs, Argo~\cite{kasapaki2016argo} provides energy-efficient, time-predictable communication using header-carried source routing while retaining routing information in packets. 
However, to route each packet, its header parsing unit decodes the route information and generates a one-hot output selection for each input, which does not support native multicast branching across multiple output ports. 
Serving multiple destinations consequently requires separate unicast transfers, introducing additional traffic and energy costs for spike fan-out. 
ChronoSpike~\cite{roshanshah2026chronospike} supports deterministic circuit-switched multicast for neuromorphic traffic while retaining per-timeslot switching tables. 
In contrast, ChroNoC supports native in-network replication without header-carried route fields or programmable per-timeslot switching tables. 

\begin{figure}[t]
    \centering
    \includegraphics[width=0.8\linewidth]{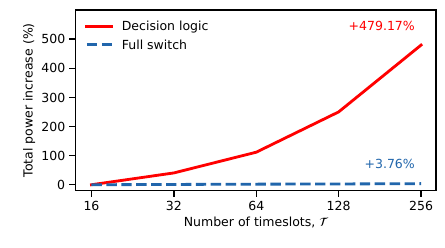}
    \caption{Relative post-synthesis power of the isolated route-decision unit and the complete switch across schedule depths $\mathcal{T}$.}
    \label{fig:power_scalability}
\end{figure}

To assess schedule scalability, the combinational route-decision unit and the complete switch were synthesized separately for $\mathcal{T}\in\{16,32,64,128,256\}$.
As shown in Fig.~\ref{fig:power_scalability}, increasing $\mathcal{T}$ from 16 to 256 raises isolated decision-unit power by 479.17\% (from 5.15 to 29.82\,$\mu$W).
Over the same range, complete-switch power increases by only 3.76\% (from 1.344 to 1.394\,mW).
Consequently, a $16\times$ increase in schedule depth incurs a modest overhead.
This low sensitivity stems from a register-dominated power profile in which sequential cells account for 81--83\% of total switch power, leaving the combinational decision logic as a minor fraction of the overall budget.
Furthermore, synthesizing combinational decoders instead of routing memories enables logic optimization across schedule depths, avoiding steep table-scaling penalties.

\subsection{Comparison with State-of-the-Art Deployment Frameworks}
\label{subsec:sota_comparison}

\begin{figure*}[t]
    \centering
    \includegraphics[width=\linewidth]{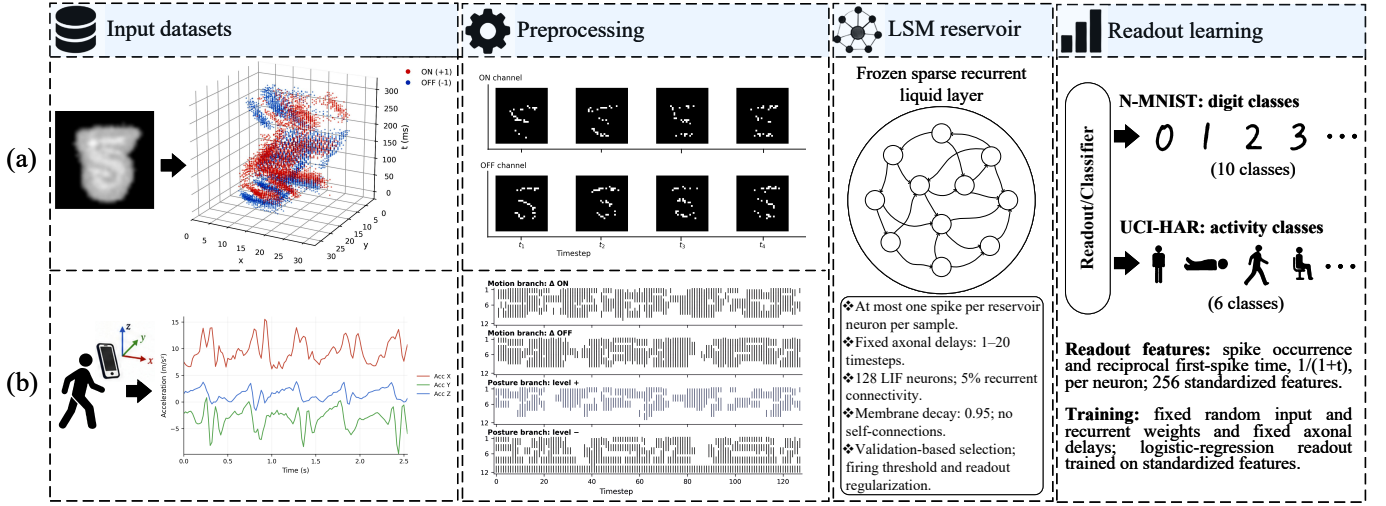}
    \caption{Workload preparation and classification pipeline for (a) N-MNIST and (b) UCI-HAR.}
    \label{fig:workload_pipeline}
\end{figure*}

Two time-sensitive neuromorphic workloads are prepared from edge benchmarks covering human activity recognition (UCI-HAR~\cite{anguita2013public}) and event-based vision (N-MNIST~\cite{orchard2015converting}).
For UCI-HAR, the nine inertial-sensor channels are augmented with three gravity components obtained by subtracting body acceleration from total acceleration. Hybrid delta and signed-level encoding converts these signals into 48 input spike channels, representing temporal changes and absolute signal levels.
For N-MNIST, positive- and negative-polarity events are aggregated into binary input frames with 1\,ms resolution.
Figure~\ref{fig:workload_pipeline} illustrates the preprocessing, recurrent reservoir, and readout training pipeline used separately for each workload.

Each encoded input stream drives a 128-neuron recurrent Leaky Integrate-and-Fire (LIF) reservoir with fixed random input and recurrent weights, a recurrent connection probability of 5\%, no self-connections, and a membrane decay factor of 0.95.
Each reservoir neuron is assigned a fixed random integer axonal delay between 1 and 20 timesteps, shared by all its outgoing connections.
During both training and inference, neurons are latched after their first firing event, limiting reservoir activity to at most 128 spikes per input sample.
Only the logistic-regression readout is trained, using 256 standardized features that encode each neuron's firing occurrence and reciprocal first-spike time, $1/(1+t)$, with zero assigned to nonfiring neurons.
The firing threshold and readout regularization parameter are selected by validation accuracy.

The recurrent connectivity provides a communication workload with cycles and multicast fan-out, while single-spike execution bounds source event counts and fixed axonal delays define delivery deadlines.
The same trained models, delays, and input samples are used across deployment methods, providing a common basis for evaluating communication cost.
Samples are processed sequentially, with all delayed spikes consumed before the reservoir is reset for the next sample.
Deployment is checked against reference neural execution to verify preservation of spike trains, neural states, and predictions; Table~\ref{tab:workloads} summarizes the workload characteristics.

\begin{table}[t]
    \centering
    \caption{Workload characteristics and parameters.} 
    \label{tab:workloads}
    \setlength{\tabcolsep}{4pt}
    \begin{tabular}{@{}lcc@{}}
        \toprule
        Property & UCI-HAR & N-MNIST \\ 
        \midrule
        Neurons / transmitting sources & 128 / 128 & 128 / 128 \\ 
        Recurrent connections & 795 & 778 \\ 
        Observation timesteps (timewindow) & 128--155 & 297--337 \\ 
        Axonal delay range & 1--20 & 1--20 \\ 
        Tasks per core (capacity) & 16 & 16 \\ 
        Test examples & 2,947 & 10,000 \\ 
        Held-out test accuracy (\%) & 75.94 & 86.55 \\ 
        \bottomrule
    \end{tabular}
\end{table}

\begin{figure*}[t]
    \centering
    \includegraphics[width=\linewidth]{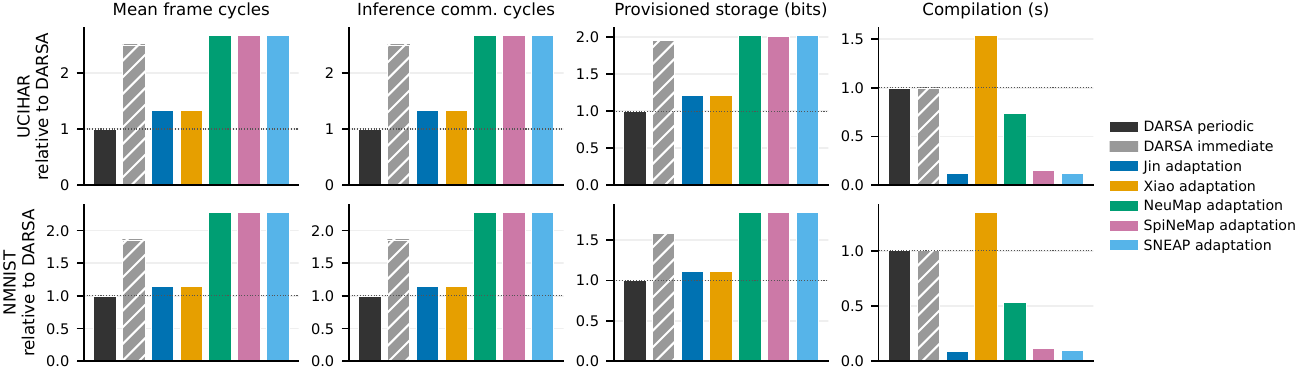}
    \caption{Deployment comparison on a $4\times4$ TDM substrate ($\mathcal{T}=59$), normalized to DARSA periodic (lower is better). Counterparts and DARSA immediate use $P_v=1$.} 
    \label{fig:deployment_compare_all}
\end{figure*}

DARSA is compared against adaptations of SNEAP~\cite{li2020sneap}, SpiNeMap~\cite{balaji2020spinemap}, NeuMap~\cite{neumap}, Xiao~\cite{xiao2024hm}, and Jin~\cite{jin2025alp} on a shared $4\times4$ substrate ($\mathcal{T}=59$), as shown in Fig.~\ref{fig:deployment_compare_all}. 
Counterparts use immediate reservations ($P_v=1$). 
DARSA immediate applies $P_v=1$ to the DARSA placement to isolate periodic reservation gains. 

Each communication frame requires $\mathcal{T}\omega$ clock cycles, excluding neuron computation and external I/O. Mean inference communication cycles over $n_{\mathrm{test}}$ examples are calculated as follows: 
\begin{equation}
    \overline{L}_{\mathrm{inference}}
    =
    \frac{1}{n_{\mathrm{test}}}
    \sum_{i=1}^{n_{\mathrm{test}}}
    H_i\mathcal{T}\omega,
    \label{eq:mean_inference_communication}
\end{equation} 
where $H_i$ also includes idle frames and the delayed-event tail. 
Relative to DARSA immediate, periodic reservations reduce $\omega$ from 15 to 6 on UCI-HAR and from 13 to 7 on N-MNIST. Frame length consequently decreases from 885 to 354 clock cycles on UCI-HAR ($60.0\%$) and from 767 to 413 clock cycles on N-MNIST ($46.2\%$). With unchanged frame counts $H_i$, mean inference communication cycles decrease by the same percentages. 
Provisioned storage (bits) sums source/destination retention (20-bit entries), NI registers (32-bit datagrams), calendar metadata, and network buffers. 
Relative to DARSA immediate, total provision drops from 94,368 to 48,396 bits on UCI-HAR and from 83,668 to 53,046 bits on N-MNIST. 
Compilation (s) requires 8.20\,s on UCI-HAR and 8.35\,s on N-MNIST (reused by DARSA immediate), trading moderate offline runtime for reduced frame duration and memory. 

\section{Conclusion}
\label{sec:conclusion}

ChroNoC and DARSA provide a hardware-software co-design for predictable communication in single-spike SNNs with prescribed axonal delays. 
ChroNoC replaces per-timeslot switch configuration memories with synthesized decoding of fixed broadcast trees. 
Operating on this fixed routing substrate, DARSA reduces deployment to a spatial search that exploits axonal delays to stagger periodic injections without advance firing-time knowledge. 
Experimental evaluations confirm reductions in per-hop energy, frame duration, and provisioned communication memory while preserving exact logical consumption timing.

\end{document}